# Efficient Password-Typed Key Agreement Scheme

Sattar J Aboud

**Information Technology Advisor**
**Baghdad-Iraq**

**Abstract**
In this paper, we will study Lee, Kim and Yoo, a verifier password typed key agreement scheme and demonstrate that the scheme is not secure. Then, the authors will propose an enhanced verifier typed key agreement scheme relied on Lee, Kim and Yoo scheme and demonstrate that the propose scheme resists against password guessing attack and stolen verifier attack. The authors are claimed that the proposed scheme is more secure and efficient compare with Lee, Kim and Yoo.

## 1. Introduction

Password-based authenticated key agreement has recently received high attention because it plays an important role to employ an authenticated key agreement schemes. In general, classical password schemes require only a human memorable shared between the entities. However, password-based key agreement schemes are prone to password guessing attacks, because the user often select the password so that it can be easily memorized, which means that they are likely to be much easier to guess than randomly selected passwords.

The key agreement protocol is one of the important techniques in public key cryptography similar to encryption and digital signature schemes. Such scheme allow two or more entities to exchange information over insecure channel and agree on a shared session key, which can employed later for secure communication between the entities. Therefore, secure key agreement scheme serve as basic building block for constructing high level of security. Secure communication requires only trusted entities that have a copy of secret key, while secret key can guarantee confidentiality, user authentication, and message integrity. In network we should able to securely distribute keys over a distance at a timely manner. It seems that, key distribution is the main problem and should be as hard as the cryptography scheme and should be able to ensure that only trusted entities have the copy of the secret key. Most schemes have an aim is to construct key on every scheme execution. In this case, the keying information define static key which will result each time the scheme is implemented by a given pair of entities. Schemes involving such fixed keys are insecure under known-key attacks.

On the other hand, dynamic key protocol is established by a group of entities differ on each execution. Dynamic key is also denoted as session key establishment. In this case the session key is dynamic, and it is generally intended that the scheme is invulnerable to known-key attacks [1]. It is preferable that each entity in a key establishment scheme is able to find out the correct identity of others which may gain access to the resulting key, including prevention of any illegal entity from deducing the same key. This needs identifications of both entities and the secret key [2]. However, the first two-party key agreement is the Diffie and Hellman scheme [3]. But, in fact Diffie and Hellman scheme is defenseless to man-in-middle attack because entities engaged with the scheme and have no channel to authenticate each other.

Also, password-typed scheme is vulnerable to dictionary attack since many entities tend to select memorable passwords of relatively low entropy [4]. In password-typed key agreement scheme the data depicted from the password is entirely common between the entities. In this case the hacker can get access to private messages, and then he can pretend any entity. To a key agreement scheme running in centralized approach, it is vulnerable to stolen-verifier attack, and the hacker who gets the verifier from the server will attempt to impersonate any entity by agree on a session key with the server.

Passwords can be simply guessed when entity selected his own password in document [5]. Storing message version of password on server is unsecure. This weakness is existed in all widely used schemes.

However, the suggested scheme is secure against dictionary attacks as long as we use only one time keys with server. The suggested scheme also offers great forward secrecy even when one key is revealed the following session keys will not be revealed. As we do not employ any public key infrastructure, great computational exponentiation is not needed.

## 2. Related Work

Since the innovative method that withstands the password guessing attacks was presented in 1989 by Lomas, Gong, Saltzer and Needham [6], there have been a several password-typed authenticated key agreement schemes were introduced.

In 1996, Jablon [7] proposed a scheme were security relied on heuristic arguments. Also, in 1999 Halevi and Krawczyk [8] introduced another scheme, the scheme considered as inflexible for security of password-typed authenticated scheme. However, Boyarsky in 1999 [9] improved this scheme by making it secure in multi-user environment, but, this scheme is inappropriate for situation where communication has to be established between entities those sharing a common limited-entropy password. In 2000 [10], another password -typed key exchange scheme has been suggested by Boyko, MacKenzie and Patel. This scheme is relied on two-party password-typed scheme. An enhancement for this scheme was made to multi-party setting by Bresson, Chevassut and Pointcheval [11]. The security of Bresson, Chevassut and Pointcheval scheme is based on the arbitrary oracle approach and in the ideal cipher approach.

In 2004, Lee, Kim, Kim and Yoo [12] suggested a verifiable-typed key agreement scheme. In this scheme, the entity employs a document of the password, while the server keeps as a verifier for the password. Thus the scheme cannot let an opponent who able to exchange information with the server to impersonate any entity without running the dictionary attack in the password file. But, the scheme is not protected against stolen-verifier attack as Kwon, in 2004[13] have claimed. Also, Yoon and Yoo Kin 2005 [14] proposed a two-party key agreement scheme relied on Diffie and Hellman scheme. Also, in 2006, Strangio [15] presented another two-party key agreement protocol relied also on Diffie and Hellman scheme. Both schemes are not appropriate for large networks since they cannot assume each party shares a secret password with other entity.

However, the first work that copes with off-line dictionary attacks is introduced in 2007 by Bellovin and Merritt [16]. They presented a family of encrypted key exchange to resist dictionary attack. This protocol is very important and become the foundation for future work in this area. In 2008, Shakir Hussain and Hussein Al-Bahadili [17] proposed simple authenticated key agreement protocol which is relied on Diffie and Hellman key agreement protocol. Unfortunately, this protocol is inefficient for practical use and does not allow concurrent executions. Also, this scheme is simple and cost effective. In 2009, SeongHan Shin, Kazukuni Kobara and Hideki Imai [18] introduced a scheme relied on threshold anonymous scheme. However, the scheme is complicated and costly.

In this paper, we will briefly evaluate Lee, Kim, Kim and Yoo 2004 [12] key agreement scheme and show its weaknesses to stolen-verifier attack. Then, we introduce a new scheme that verifier-typed key agreement scheme. The new scheme resists password guessing attack and stolen-verifier attack.

## 3. Lee, Kim and Yoo Scheme

In 2004 Lee, Kim and Yoo [12] introduced a verifier based key agreement scheme. They claimed that the proposed scheme was secure in the case of server compromise. It indicates that when the hacker attacks the server, he cannot obtain sufficient information to pose as an entity without execution a dictionary attack on the password file. Now, we briefly describe their scheme which is as follows:

3.1 Notations Used
$A$ : Entity communicate with entity $B$ the sever
$B$ : Server entity
$v$ : Verifier computed by entity $A$
$P$ : The password
$q$ : Secure large prime number
$g$ : Generator in $Z_q^*$ of order $q-1$.
$id_A$ : Identification of entity $A$
$id_B$ : Identification of entity $B$
$x, y$ : Two integer randomly selected numbers of order $Z_q^*$.
$h$ : Secure one-way hash function.
$\oplus$ : The XOR function

3.2 The Scheme Description
Suppose that there is a secure one way hash function $h : \{0,1\}^* \to Z_q^*$. The steps of the scheme are as follows:

Step 1: Entity $A$
  1.1. Select a password $P$
  1.2. Find $v = g^{h(id_A, id_B, P)}$
  1.3. Pass $v$ to the server entity $B$ as the Verifier.
  1.4. Select an arbitrary number $x \in Z_q^*$
  1.5. Find $T_A = g^x \oplus v$

1.6. Pass $(id_A, T_A)$ to server entity $B$

**Step 2: Entity** $B$
2.1. Select an arbitrary number $y \in Z_q^*$
2.2. Find $T_B = v^y \oplus v$
2.3. Find $r_B = (T_A \oplus v)^y = g^{x*y}$
2.4. Compute $d_A^{'} = h(id_A, T_B, r_B)$
2.5. Find $d_B = h(id_B, T_A, r_A)$
2.6. Pass $T_B$ and $d_B$ to entity $A$

**Step 3: Entity** $A$
3.1. Find $r_A = (T_B \oplus v)^{x*h(id_A, id_B, P^{-1})} = g^{x*y}$)
3.2. Compute $d_A = h(id_A, T_B, r_A)$
3.3. Compute $d_B = h(id_B, T_A, r_A)$
3.4. pass $d_A$ to entity $B$

**Step 4: Entity** $B$
4.1. Verify if $d_A = d_A^{'}$ entity $B$ authenticates entity $A$
4.2. Find the common session key $r = h(r_A) = h(g^{x*y})$.

**Step 5: Entity** $A$
5.1. Verify $d_B = d_B^{'}$ entity $A$ authenticates entity $B$
5.2. Find the common session key $r = h(r_B) = h(g^{x*y})$.

### 3.3 Vulnerabilities

Lee, Kim and Yoo claimed that the scheme was secure in the case of server compromise. But, in 2005 Shim and Seo [19] stated that the scheme was weak against stolen verifier attack. On the other hand, given the verifier, the hacker can impersonate an authorized entity $A$ to negotiate a session key with the server entity $B$. The weakness of the scheme is that entity $B$ has not an efficient way to verify the message claimed to be sent by entity $A$. So, we develop the scheme of Lee, Kim and Yoo by introducing a new verifier-typed authenticated protocol, which resists against stolen verifier attack.

## 4. The Proposed Password Scheme

The description of the scheme is as follows:

### 4.1 Algorithm of the Proposed Scheme
The steps of the algorithm are as follows:

**Step 1: Entity** $A$
1.1. Select a password $P$
1.2. Find $v = g^{h(id_A, id_B, P)}$
1.3. Pass $v$ to the server entity $B$ as the Verifier.
1.4. Select an arbitrary number $x \in Z_q^*$
1.5. Find $T_A = g^x \mod q$
1.6. Pass $T_A$ to server entity $B$

**Step 2: Entity** $B$
2.1. Select an arbitrary number $y \in Z_q^*$
2.2. Find $T_B = v^y \mod q$
2.3. Pass $T_B$ and to entity $A$

**Step 3: Entity** $A$
3.1. Find $r = (T_B)^{x*h(id_A, id_B, P)^{-1}} \mod q$
3.2. Compute $d_A = h(r) \mod q$
3.3. Pass $d_A$ to entity $B$

**Step 4: Entity** $B$
4.1. Find $F_A = h(T_A)^y \mod q$
4.2. Verify if $F_A = d_A$ entity $B$ authenticates entity $A$
4.3. Compute $E_B = v^{y^2} \mod q$
4.4. Find the common session key $r = (h(id_A, id_B, T_A^y)$

**Step 5: Entity** $A$
5.1. Verify $e(E_B, g) = e(T_B, T_B^{x(id_A, id_B, p)^{-1}}) \mod q$ if yes, entity $A$ authenticates entity $B$
5.2. Find the common session key $r = h(id_A, id_B, T_B^{x*h(id_A, id_B, P)^{-1}}) \mod q$.

Upon successfully implementing above scheme the two entities, will agree on a shared session key $r = h(id_A, id_B, g^{x*y})$.

**Example**
Suppose that the prim $q = 13, g = 6, id_A = 9, id_B = 12, P = 10$
**Step 1: Entity** $A$
$v = (g = 6^{h(id_A=9, id_B=12, P=10)}) \mod 13$
$= 6^{(9+12+10)} \mod 13 = 7$, then pass this result to entity $B$
Suppose $x = 3$
$\therefore T_A = 6^{x=3} \mod 13 = 8$, send this value to entity $B$
**Step 2: Entity** $B$
Suppose $y = 4$
$\therefore T_B = 7^{y=4} \mod 13 = 9$, and send this value to entity $A$
**Step 3: Entity** $A$
Find $r = (9)^{3*h(9,12,10)^{-1}} \mod q$
$= 9^{3(31)^{-1}} \mod 13 = 9^{3(8)} \mod 13 = 9^{24} \mod 13 = 1$
Compute $d_A = h(1) \mod q = 1$, send this value to entity $B$
**Step 4: Entity** $B$
4.1. Find $F_A = h(8)^4 \mod q = 1$

4.2. Verify if $F_A = d_A$ entity $B$ authenticates entity $A$

4.3. Compute $E_B = v^{y^2} \mod q = 7^{4^2} \mod 13 = 9$

4.4. Find the common session key
$r = (h(id_A, id_B, T_A^y) \mod q = (9, 12, 8^4) = 4117 \mod 13 = 9$

### 4.2 Security Discussions

We will show that the proposed password typed key agreement protocol is secure against both password guessing attack and stolen verifier attack.

1. **Resist Man-in-Middle Attack**: The pre-shared password and verifier are employed to stop the man-in-middle attack is easy because a hacker does not have the verifier or password; it means that the hacker cannot impersonate entity $A$ to exchange information with entity $B$.

2. **Resist Dictionary Attack**: To the on-line password guessing attack, the entities can overcome the hacker by selecting suitable trail intervals. In an off-line guessing attack, the hacker must repeatedly guess the password and check its accuracy by the message collected in an off-line approach. In the proposed scheme, the hacker is allowed to gather any message exchanged through the channel. It means that the hacker can get $g^x, g^{y*x}, h(g^{x*y})$, $g^{y^2*x}$ since $x, y \in Z_q^*$ are arbitrary numbers uniformly distributed in $Z_q^*$ the off-line dictionary attack is beaten. In addition, known $g^{y*x}$ and $g^{x^2*x}$ a hacker cannot obtain $g^y$ by the proposed scenario. As a result, we can mention that the suggested scheme is secure against dictionary attack.

3. **Resist Stolen Verifier Attack**: Suppose that a hacker, entity has imposed entity $B$ and obtained the verifier. A hacker goal is to impersonate entity $A$ to negotiate a session key with entity $B$. We have the following theorem.

**Theorem**: Assume that we have the key agreement protocol is secure against stolen verifier attack.

**Proof**: In this scenario, hacker is allowed to select an arbitrary number $x \in Z_q^*$ and finds $T_A = g^x$. We assume that the hacker has aptitude to impersonate entity $A$. On the other hand, hacker must produce two results $T_A$ and $d_A$ which satisfy $r_A = d_A$

As $h$ is a robust one way hash function, obtained $g^x$ and $g$, hacker must calculate $g^y$ and then utilize this result to calculate $d_A$ hence verifier $r$ is indicated by $g^x$.

Clearly, it is different from the complexity scenario illustrated previously. There is an alternative technique for hacker to impersonate entity $A$. Hacker can gather messages $g^x, g^{y*x}$ and $g^{y^2*x}$ then attempt with the obtained result. But, the scenario described previously is intractable. From depicted above we can summarize that the hacker cannot impersonate entity $A$ even if he gets the verifier kept in Server and attempts to make stolen verifier attack.

### 4.3 Efficiency

Efficiency of the proposed protocol is related to the costs of communication and computation. Communication cost involves counting total number of rounds and total messages transmitted through the network during a protocol execution. Number of rounds is a critical concern in practical environments where number of group members is large. Compares the proposed protocol with Lee, Kim and Yoo password typed key agreement protocol.

Concerning cost communications, the suggested protocol requires only two rounds while Kim and Yoo require $n$ rounds; where every user sends one message in every round. Regarding the maximum bit length of messages sent per user during the execution of the proposed protocol is $2|e|$ such that $|e|$ is the maximum size of an encrypted message compare with $n|e|$ in Lee, Kim and Yoo password typed key agreement protocol. Concerning the maximum number of point-to-point communication per user, the proposed protocol require $n+1$ while Lee, Kim and Yoo password typed key agreement protocol require $2n-2$. To understand this case consider the users $U_1, ..., U_n$ participating in the protocol are on a ring and $U_{i-1}, U_{i+1}$ are respectively the left and right neighbors of $U_i$ for $1 \le i \le n$ such that $U_0 = U_n, u_{n+1} = U_1$. User $U_i$ where $1 \le i \le n-1$, sends a message in round 1 only to the users $U_{i-1}, U_{i+1}$ and a message in round 2 to the rest of the $n-1$ users whilst the last user $U_n$ sends one message in each round to all the $n-1$ users. These will make the proposed protocol efficient from communication viewpoint.

Regarding cost computation, in the proposed protocol every group member executes at most 3 modular exponentiations compared with $2n$ in Lee, Kim and Yoo protocol. Also, the proposed protocol requires 4 one-way hash function evaluations, 2 encryptions and $n+1$ decryption operations. The operations dependent on the number of group members are the asymmetric key decryption operation, compared with 1 encryption and 2 decryptions in Lee, Kim and Yoo protocol. The total cost of computation is highly reduced compared to Lee, Kim and Yoo protocol password typed key agreement protocol. We use asymmetric key encryption and decryption. Hence the proposed protocol attains efficiency in both communication and computation costs. The constant round protocol can be implemented for a large group of participants as compared to Lee, Kim and Yoo protocol password typed protocol which becomes not practical if $n > 100$.

## 5. Conclusions

In this paper, we have shown that Lee, Kim and Yoo password-typed key agreement protocol is vulnerable to the password guessing attack and stolen verifier attack. To avoid these attacks, we presented a modified verifier-typed key agreement scheme relied on Lee, Kim and Yoo protocol and demonstrate that the propose scheme resists against password guessing attack and stolen verifier attack. According to the security analysis, it is obvious that the modified protocol is secure enough to withstand all possible mentioned attacks. Constructing password schemes using authenticated key agreement has received high attention in the last decade. In practice, password-typed protocols are appropriate for implementation in many situations, especially where no device is able of securely storing high-entropy long-term secret key. As we are mentioned, password has low entropy and is vulnerable to dictionary attack and man-in-middle attack, researchers must be cautious in construction password-typed scheme.

Different channel characteristic and different environment need to be studied to determine further useful relations. Since multiple channels might increase overheads, studies might be done to consider the best environment combinations to reach high security at the least cost. Work also remains to be done to formalize these schemes

**Sattar J Aboud** received his B.S. degree in 1976. In 1982, he earned his Master degree in computing science. A Ph.D. was received in 1988 in the area of computing science. The last two degrees were awarded from U.K. In 1990, he joined the institute of technical foundation, in Iraq as an assistant professor and a head of computer system department. In 1993 he moved to Arab University College for science and technology in Iraq as an associate professor and a dean deputy. In 1995 he joined the Philadelphia University in Jordan as an associate professor and a chairman of computer science and information system department. In 2004 he moved to the Amman Arab University for graduate studies, graduate college for computing studies as a professor. Currently, he is a professor in the department of computer information systems at the Middle East University for graduate studies, Amman-Jordan. His research interests include areas like public key cryptography, digital signatures, identification and authentication, software piracy, networks security, data base security, e-commerce and e-learning security and algorithm analyzes and design. He has supervised many PhD's and master's degrees research thesis and projects of diverse areas. He has published more than 70 research papers in a multitude of international journals and conferences.
.